\documentclass{eas}
\usepackage{graphicx}
\begin{document}

\title{Micromegas detector developments for MIMAC} 

\author{E. Ferrer-Ribas}\address{CEA/DSM/IRFU,91191 Gif sur Yvette,  France}
\author{D. Atti\'e}\sameaddress{1}
\author{D. Calvet}\sameaddress{1}
\author{P. Colas}\sameaddress{1}
\author{F.~Druillole}\sameaddress{1}
\author{Y. Giomataris}\sameaddress{1}
\author{F.J. Iguaz}\sameaddress{1}
\author{J.P. Mols}\sameaddress{1}
\author{J. Pancin}\address{GANIL, Bvd H. Becquerel, 14076 Caen, France}
\author{T. Papaevangelou}\sameaddress{1}
\author{J.~Billard}\address{LPSC,  Universite Joseph Fourier Grenoble 1, CNRS/IN2P3, Institut Polytechnique de
Grenoble, France} 
\author{G.~Bosson}\sameaddress{3}
\author{J.L.~Bouly}\sameaddress{3}
\author{O.~Bourrion}\sameaddress{3}
\author{Ch.~Fourel}\sameaddress{3}
\author{C.~Grignon}\sameaddress{3}
\author{O.~Guillaudin}\sameaddress{3}
\author{F.~Mayet}\sameaddress{3}
\author{J.P.~Richer}\sameaddress{3}
\author{D.~Santos}\sameaddress{3}
\author{C.~Golabek}\address{IRSN Cadarache, 13115 Saint-Paul-Lez-Durance, France} 
\author{L.~Lebreton}\sameaddress{4} 

\begin{abstract}
The aim of the MIMAC project is to detect non-baryonic Dark Matter with a directional TPC.  The recent Micromegas efforts towards building a large size detector will be described, in particular the characterization measurements of a prototype detector of 10 $\times$ 10~cm$^2$ 
with a 2 dimensional readout plane. Track reconstruction with alpha particles will be shown.
\end{abstract}
\maketitle
\section{Introduction}

The MIMAC (MIcro TPC MAtrix of Chambers) collaboration (Santos {\em et al.\/} \cite{Santos2010}) aims at building a directional Dark Matter detector composed of a matrix of Micromegas (Giomataris {\em et al.\/} \cite{Giomataris1996}) detectors. The MIMAC project is designed to measure both 3D track and ionization energy of recoiling nuclei, thus leading to the possibility to achieve directional dark Matter detection (Spergel {\em et al.\/} \cite{Spergel1988}). It is indeed a promising search strategy of galactic Weakly Interacting Massive Particles (WIMPs) and several projects of detector are being developed for this goal (Ahlen {\em et al.\/} \cite{Ahlen2010}). Recent studies have shown that a low exposure CF$_4$ directional detector could lead either to a competitive exclusion (Billard {\em et al.\/} \cite{Billard12010}), a high significance discovery (Billard {\em et al.\/} \cite{Billard22010}), or even an identification of Dark Matter (Billard {\em et al.\/} \cite{Billard2011}), depending on the value of the WIMP-nucleon axial cross section.

Gaseous detectors are attractive as they can reconstruct the track of the nuclear recoil providing both 
the energy and the track properties. Micropattern gaseous detectors are particularly suited to reconstruct low energy (few keV) 
recoil tracks  of a few mm  length due to their very good granularity, good spatial and energy resolution and low threshold. Micromegas detectors have shown these qualities in different environments (Aune {\em et al.\/} \cite{Aune2009}; Dafni {\em et al.\/} \cite{Dafni2009}; Dafni {\em et al.\/} \cite{Dafni2011}; Delbart {\em et al.\/} \cite{Delbart2010}). In particular thanks to the new manufacturing techniques, namely bulk (Giomataris {\em et al.\/} \cite{Giomataris2006}) and microbulk (Andriamonje {\em et al.\/} \cite{Andriamonje2010}), where the amplification region is produced as a single entity. 

The concept of the MIMAC experiment is described in detail in the contribution (Santos {\em et al.\/} \cite{Santos2011}) of these proceedings. This paper is devoted to the recent Micromegas developments done with bulk detectors in order to show  the feasability of a large TPC (Time Projection Chamber) for directional detection.
\section{The 10 $\times$ 10~cm$^2$ MIMAC prototype}
\begin{figure}[htb!]
\centering%
\begin{tabular}{cc}
\includegraphics[width=0.5\textwidth]{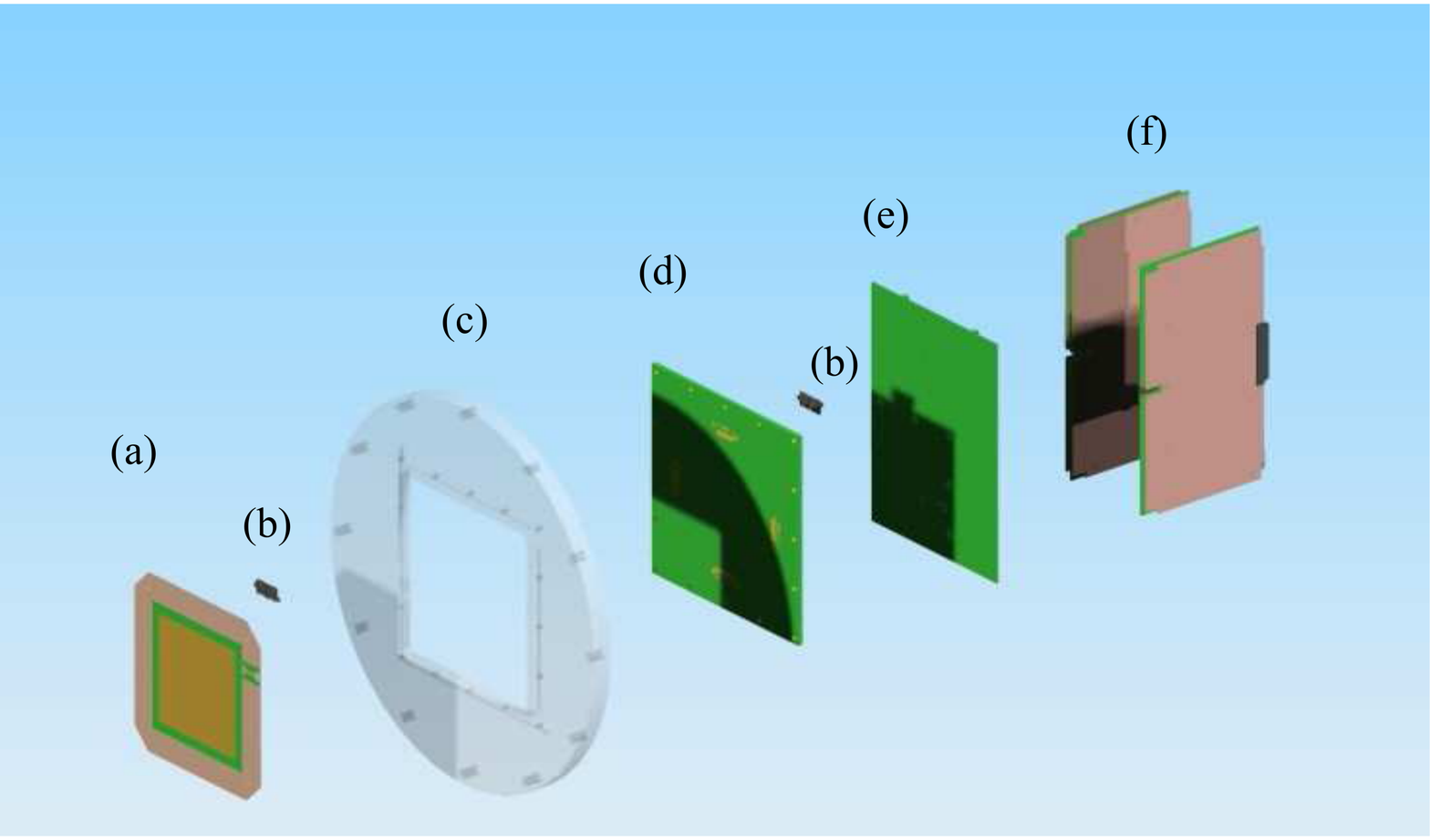} &
\includegraphics[width=0.5\textwidth]{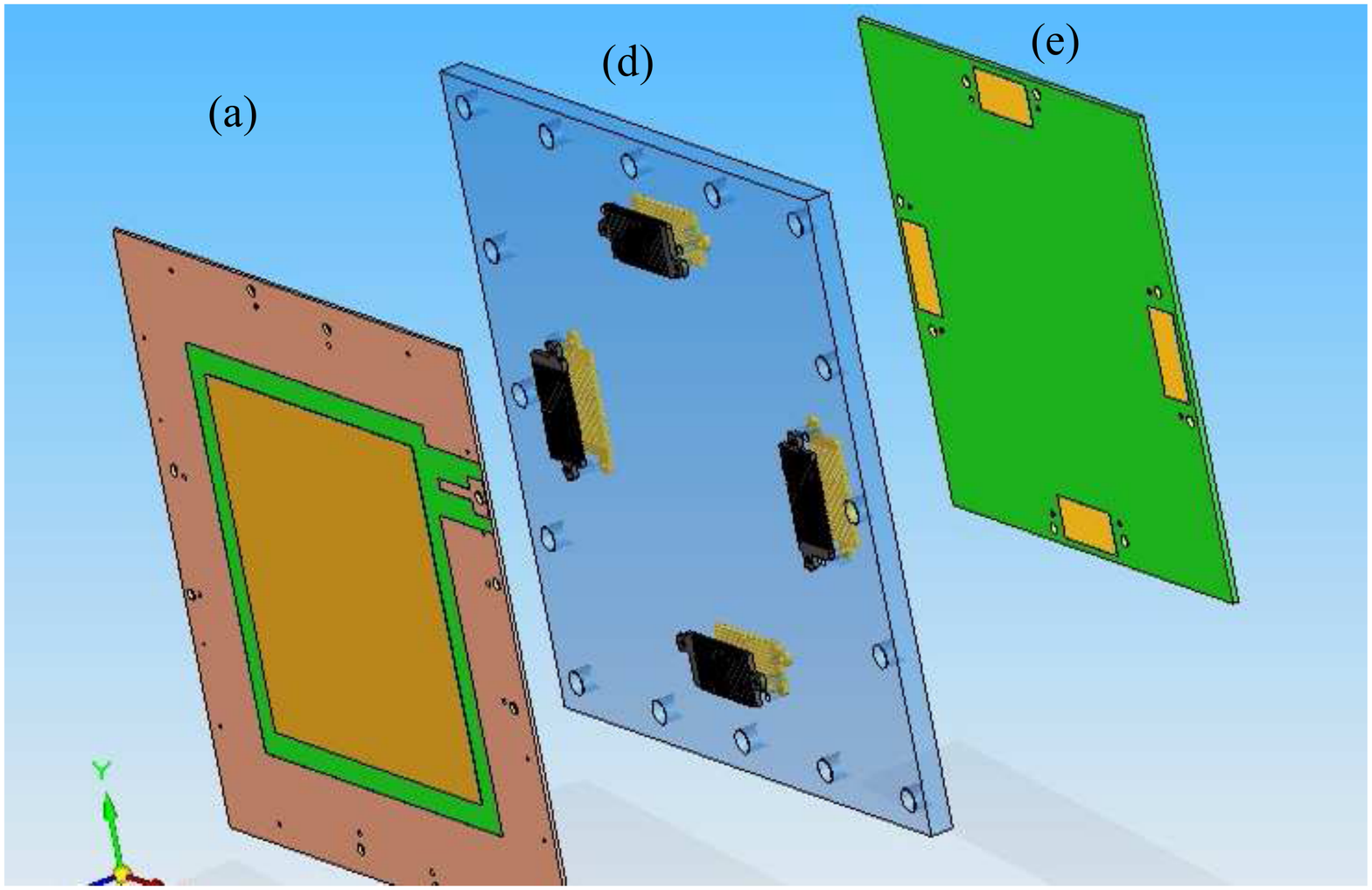}
\end{tabular}
\caption{Left: Sketch of the general assembly constituting the readout plane and the interface to electronics. Left: Zoom of the Readout PCB (\emph{a} in the image), Leak tight PCB (\emph{d}) and the Interface PCB (\emph{e}). }
\label{fig:assembly}
\end{figure}

In order to have a detector that can stand operation at low and high pressure the design relies on the idea of assembling a 
leak-tight read-out plane on a 2\,cm aluminium cap.  A general sketch of the mechanical assembly is given in figure~\ref{fig:assembly}.

\begin{figure}[htb!]
\centering%
\begin{tabular}{cc}
\includegraphics[width=0.45\textwidth]{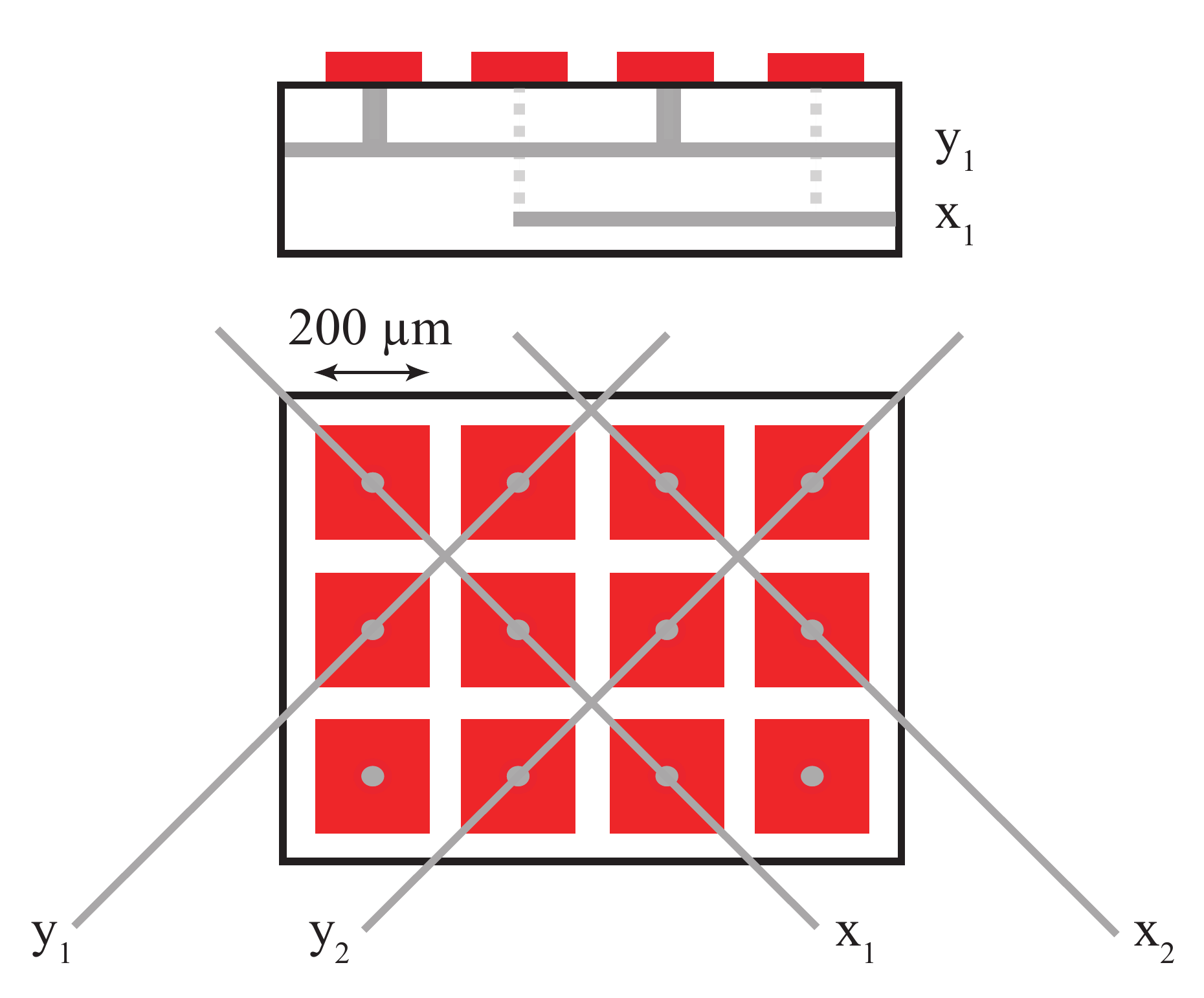} &
\includegraphics[width=0.45\textwidth]{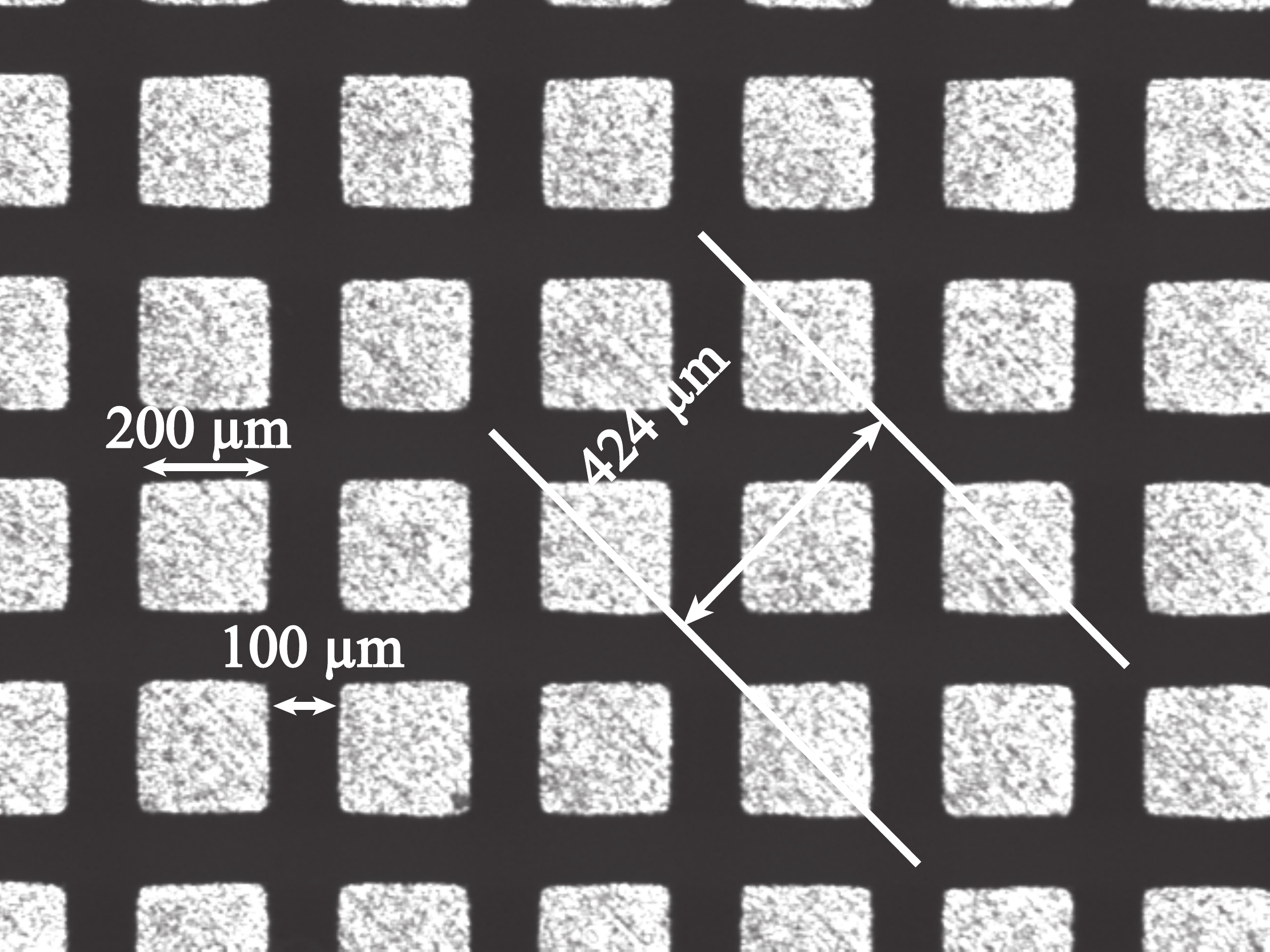}
\end{tabular}
\caption{Left: Sketch of the 2D readout used. Right: Microscope photograph of the 2D readout.}
\label{fig:2d}
\end{figure}

The bulk Micromegas is on a Printed Circuit Board (PCB), called \emph{Readout PCB}, of 1.6\,mm thickness (\emph{a} in figure~\ref{fig:assembly}).  The active surface is of 10.8$\times$ 10.8\,cm$^2$ with 256 strips per direction. The charge collection strips make-up an X--Y structure out of electrically connected pads in the diagonal direction through metallized holes as can be seen in figure~\ref{fig:2d} (left). This readout strategy  reduces the number of channels with a fine granularity covering a large anode surface. The pads are 200\,$\mu\mathrm{m}$ large with an isolation of  100~$\mu\mathrm{m}$ resulting into a strip pitch of 424~$\mu\mathrm{m}$. The quality of the surface of the readout plane can be observed in figure~\ref{fig:2d} (right). The 100\,$\mu\mathrm{m}$ diameter metallized holes have been fully filled yielding a completely uniform surface.  This fact is a prerequisite to  obtain a uniform performance of a bulk Micromegas detector.

The strips signals are rooted into 4 connectors prints  at the sides of the Readout PCB. The Readout PCB is screwed on a thick 0.5\,cm PCB, called \emph{Leak Tight PCB} (\emph{d} in figure~\ref{fig:assembly}), that will ensure the leak tightness of the system. The Leak Tight PCB is constituted of various layers of FR4 with blind metallized vias in the inner layer. This piece is then screwed on a 2\,cm thick aluminium cap that constitutes the bottom of the TPC (\emph{c} in figure~\ref{fig:assembly}). The signal connections from one board to another are done by means of SAMTEC connectors (GFZ 200 points) that are placed and screwed between the two boards (\emph{b} in figure~\ref{fig:assembly} (left)). On the outside of the vessel an \emph{Interface card} distributes the signals to the desired electronics (\emph{e} in figure~\ref{fig:assembly}). 
Two versions of this card exists: one dedicated to the laboratory set-up and a second one for the final MIMAC electronics.

This design offers several advantages: a simple, compact and leak-tight way for the signal connections and a versatility for two different types of electronics. 
Bulk Micromegas with two different amplification gaps were produced in order to choose the best gap for different running pressure conditions. Characterisations tests described in the next section concern three readout planes with a gap of 128\,$\mu$m and two with a gap of 256\,$\mu$m.

\section{Characterization measurements} 
\label{sec:characterisation}
\begin{figure}[htb!]
\centering%
\includegraphics[width=0.8\textwidth]{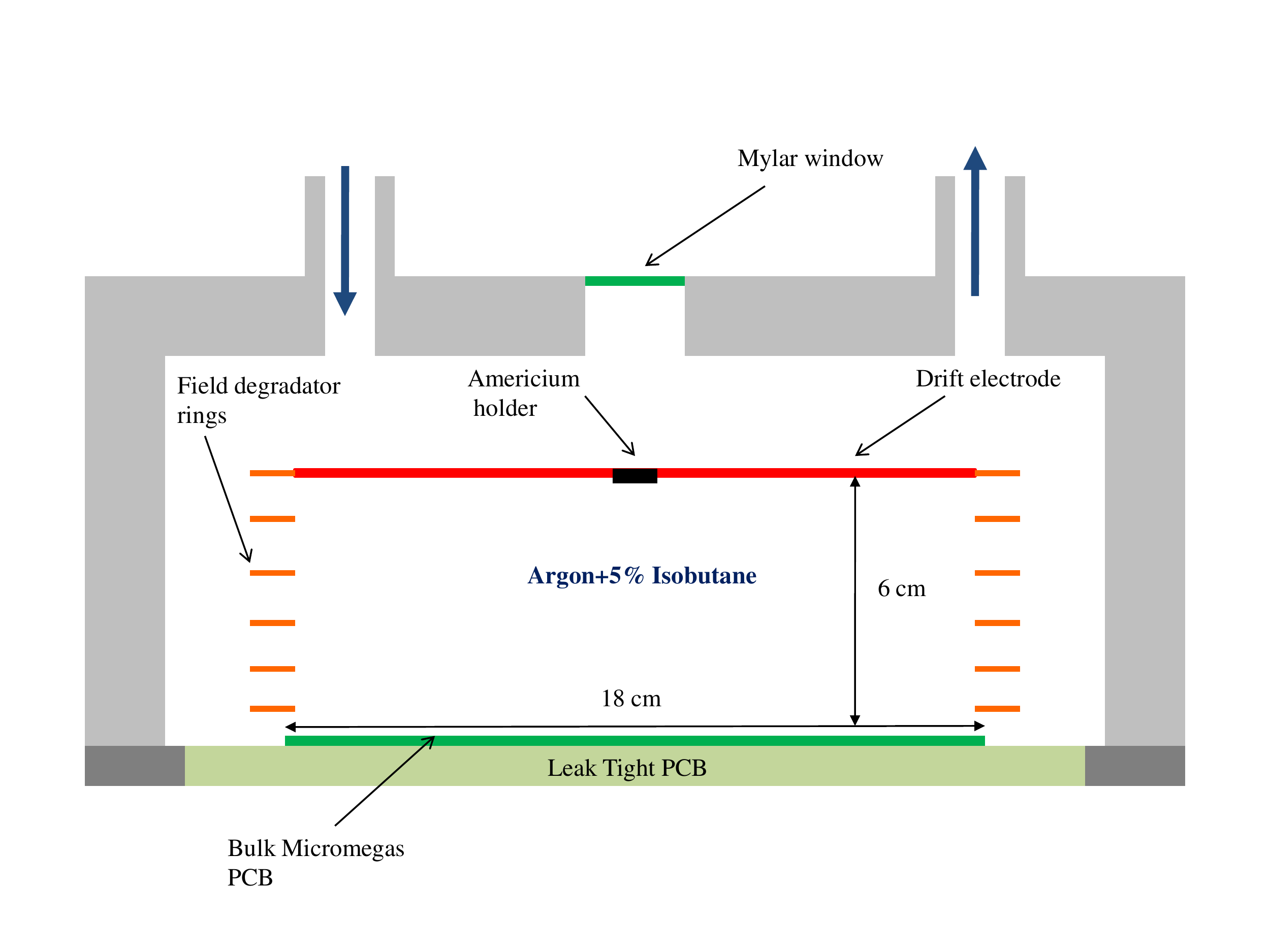} 
\caption{Schematic drawing of the set-up.}
\label{fig:setup}
\end{figure}

\begin{figure}[htb!]
\centering
\includegraphics[width=60mm]{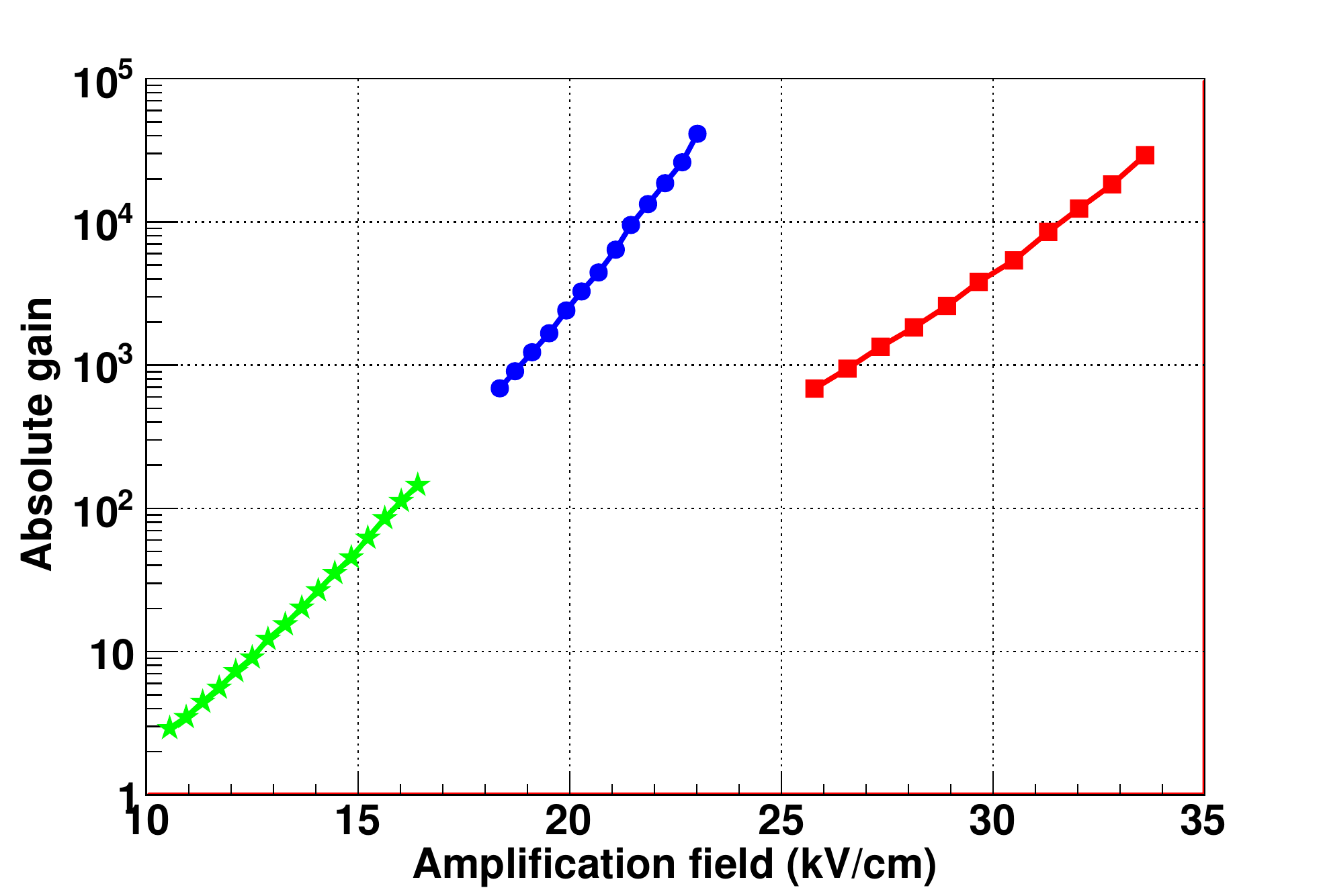}
\includegraphics[width=60mm]{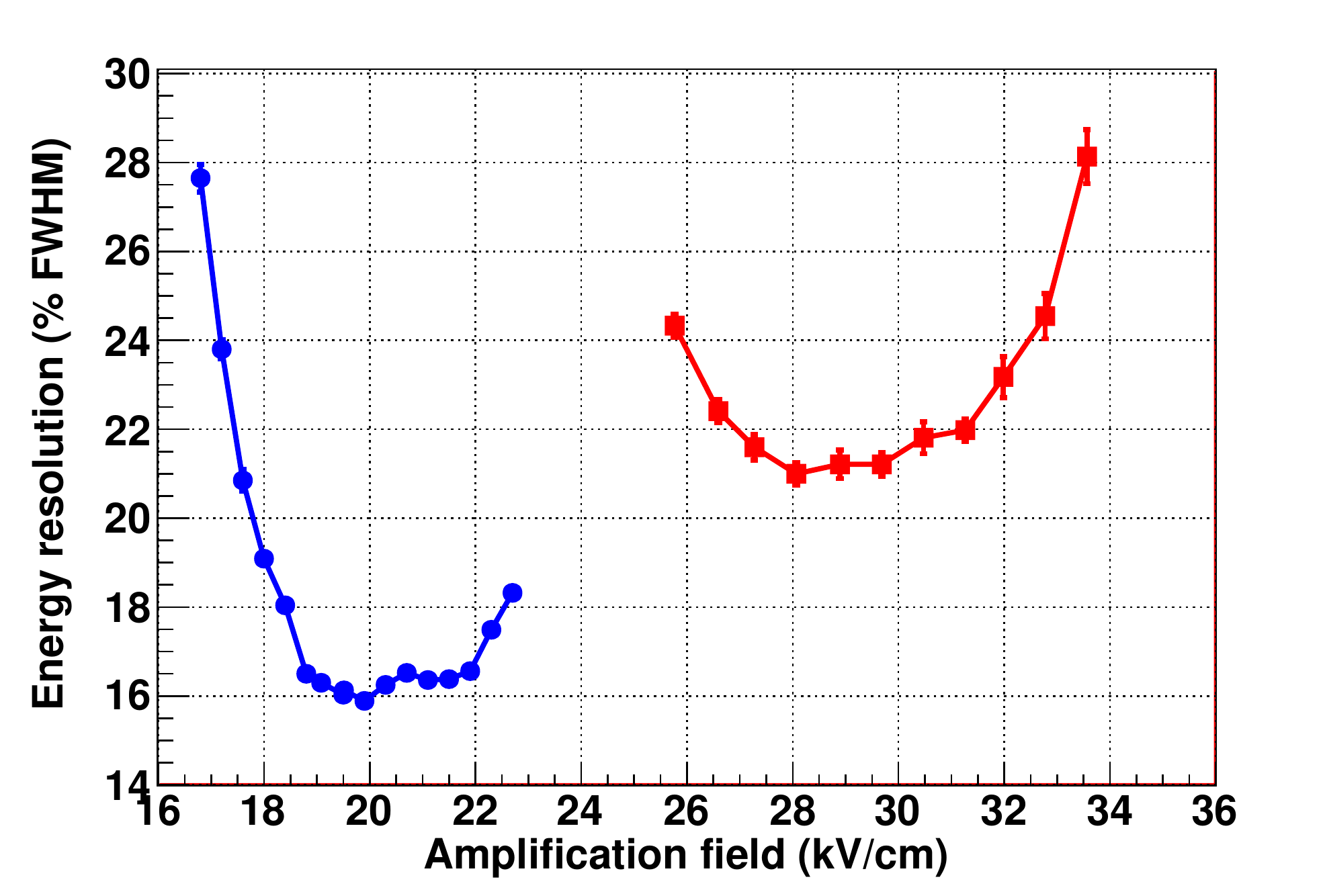}
\caption{ Left: The absolute gain as a function of the amplification field is shown for the readouts with a 128\,$\mu$m (red squared line) and 256\,$\mu$m-thick amplification gap (blue circled line) obtained with a iron $^{55}$Fe source using an Argon(95\%)-Isobutane(5\%) mixture at 1\,bar. The extension at low gain (green star line) for the
256\,$\mu$m readout was obtained with an alpha source. The maximum gain was reached just before the spark limit.  
Right: Dependence of the energy resolution at 5.9 keV with the amplification field for the readouts of 128$\mu$m (red squared line) and 256 $\mu$m-thick amplification gaps (blue circle line).}
\label{fig:GainRes}
\end{figure}

A dedicated vessel was built as shown in figure~\ref{fig:setup}. The set-up is described in detail in 
(F.J Iguaz {\em et al.\/} \cite{MimacSaclay}).
The characterization measurements were realized in an Argon(95\%)-Isobutane(5\%) mixture at atmospheric pressure. The strips were connected to ground and the readout was illuminated by an iron $^{55}$Fe source (X-rays of 5.9 keV) located on the TPC window.  The mesh signal  was read out by an ORTEC 142C preamplifier, whose output was fed into an ORTEC 472A Spectroscopy amplifier and subsequently into a multi-channel analyzer AMPTEK MCA-8000A for spectra building. In order to measure the gain of the detectors the ratio of drift and amplification fields was fixed in the region where the mesh showed the maximum electron transmission. The tested readouts  reach gains greater than $2 \times 10^4$ before the spark limit for both amplification gaps as can be seen in figure~\ref{fig:GainRes} (left). The energy resolution stays constant for a wide range of amplification fields as shown in figure \ref{fig:GainRes} (right). At low fields, the resolution worsens due to the noise level that is comparable to the signal height. At high fields, it deteriorates due to the gain fluctuations. 

\section{Track measurements}
\begin{figure}[htb!]
\centering
\includegraphics[width=60mm]{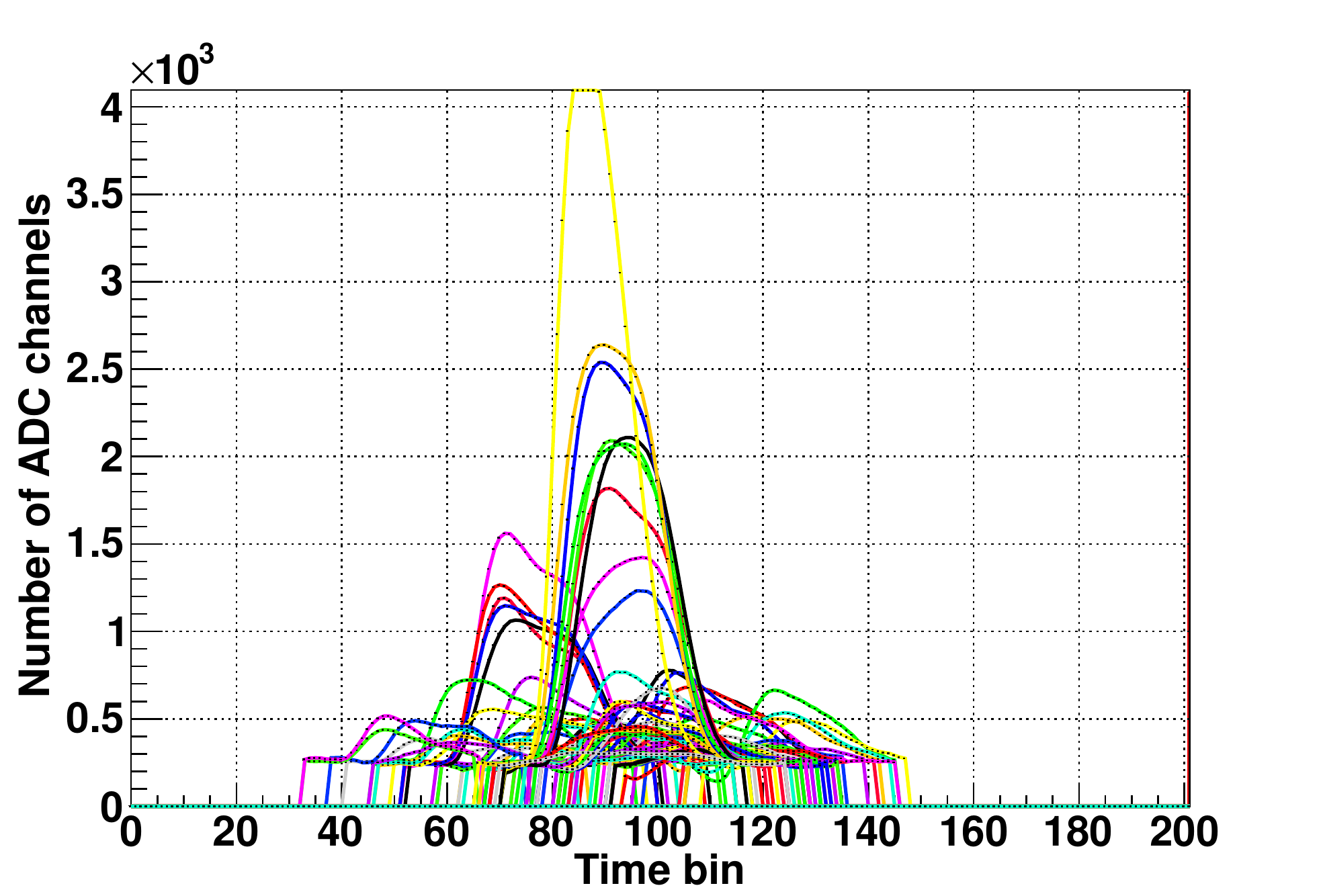}
\includegraphics[width=60mm]{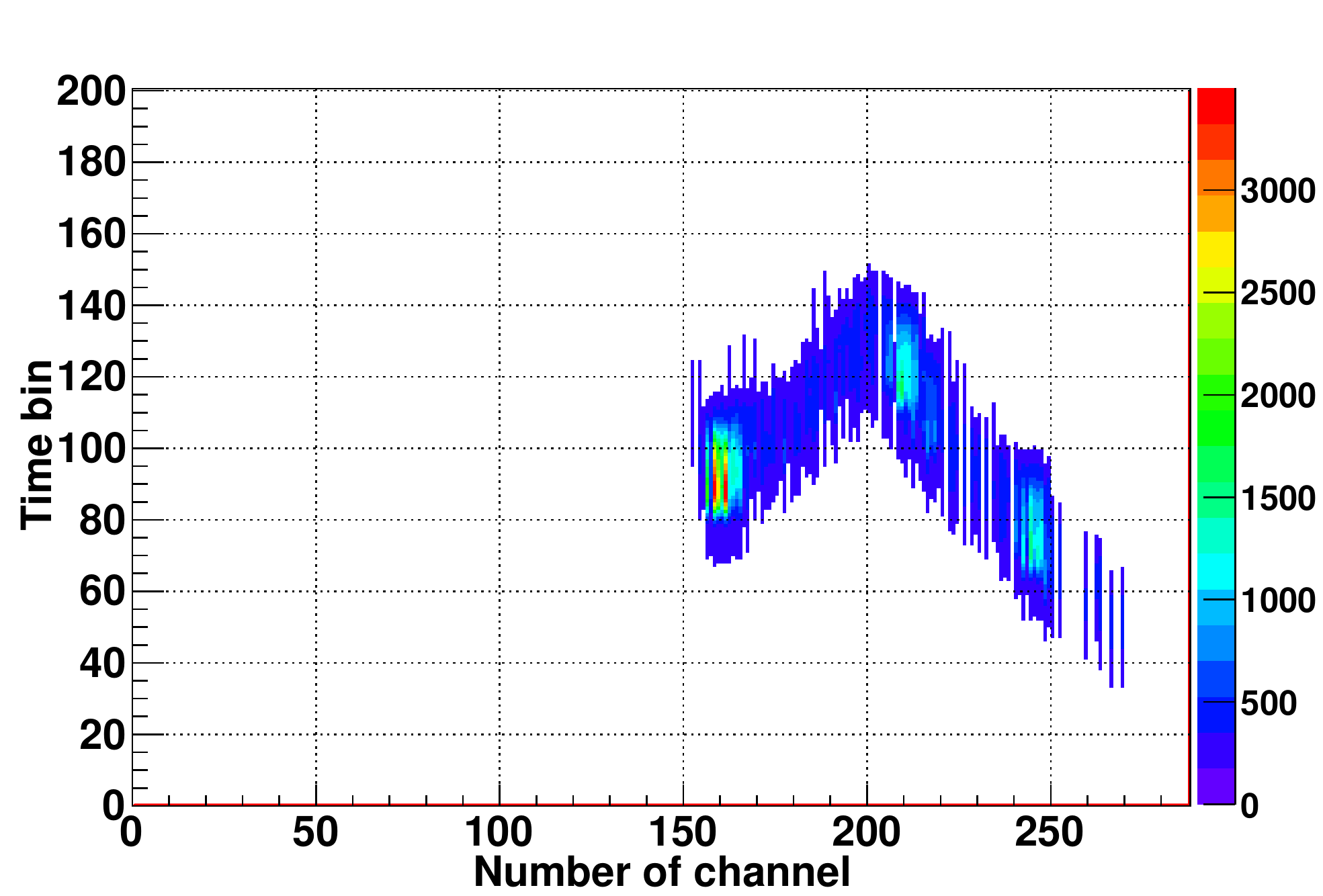}
\caption{ Left: Example of pulses induced in the strips acquired with the T2K electronics. Right: The reconstruction of the XZ projection of the same event. The physical event is an electron candidate of 44.4\,keV with a final charge accumulation (or blob).}
\label{fig:PulseEvent}
\end{figure}

The T2K electronics Baron {\em et al.\/} (\cite{Baron2009}, \cite{Baron2010}) has been used to read the signals induced in the strips to fully validate the concept of MIMAC readouts for the reconstruction of tracks. An example of the strips pulses and the XZ reconstruction of one event is shown in figure \ref{fig:PulseEvent}.

In the two spatial projections (x and y) of the event, different parameters characterising the charge were calculated like mean position, width and number of activated strips. The analysis was then extended to the perpendicular direction using the amplitudes of strips pulse in each temporal bin. Finally, the total charge of each event was obtained summing the charge of both projections. 
After having tested the readout with low energy events, we evaluated its performance at low gain with high energy events. For this purpose, an $^{241}$Am alpha source was installed in the source keeper screwed at the center of the drift plate. The source consists of a metallic circular substrate of 25\,mm diameter where the radioactive material has been deposited on its center in a circular region of about 8\,mm of diameter. The alpha particles are emitted isotropically. The 5.5 MeV\,alphas were used to characterize the readout as it was done with the iron source in section \ref{sec:characterisation}, generating the gain curves shown in figure \ref{fig:GainRes} (left).

\begin{figure}[htb!]
\centering
\includegraphics[width=60mm]{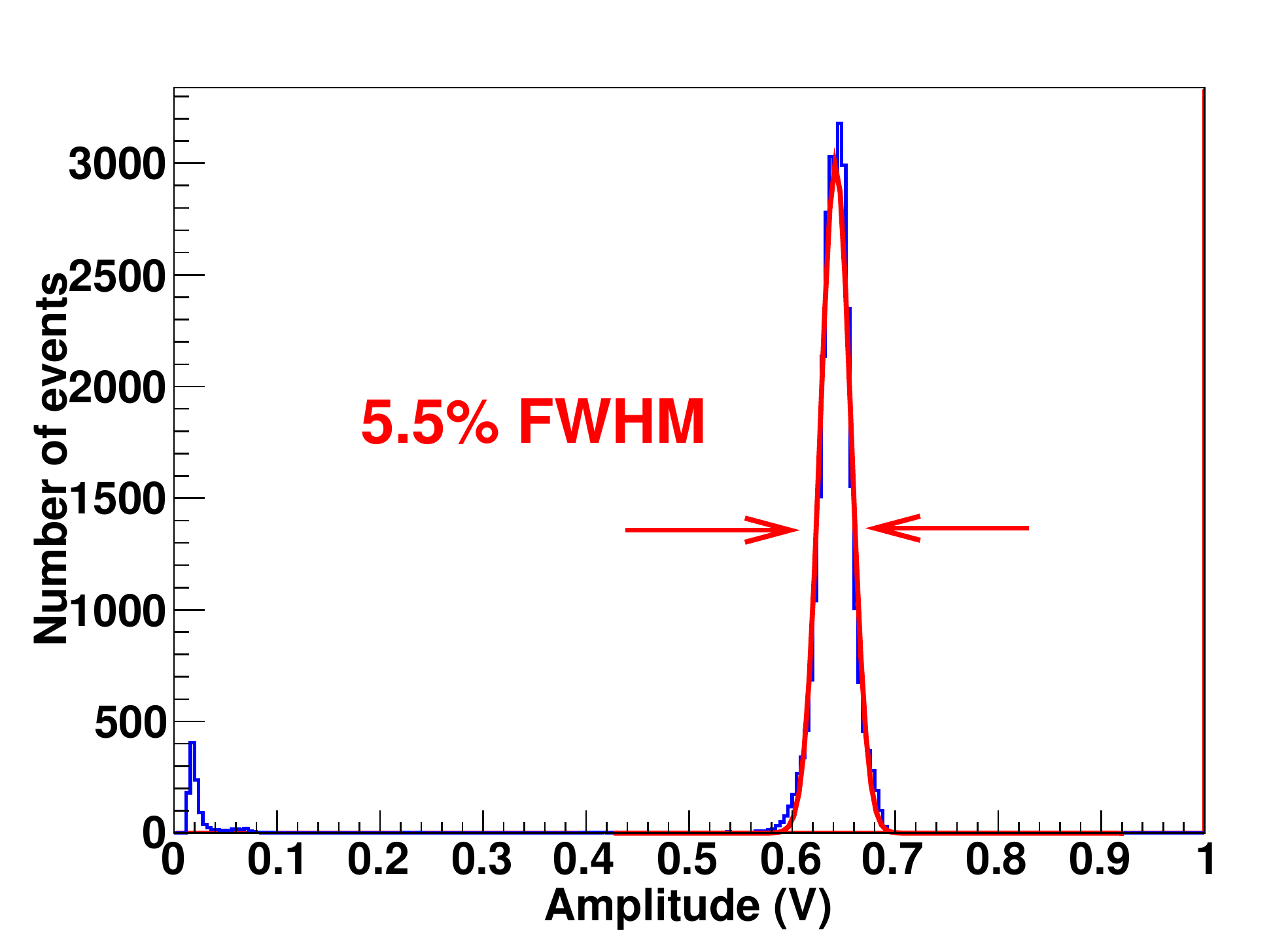}
\includegraphics[width=60mm]{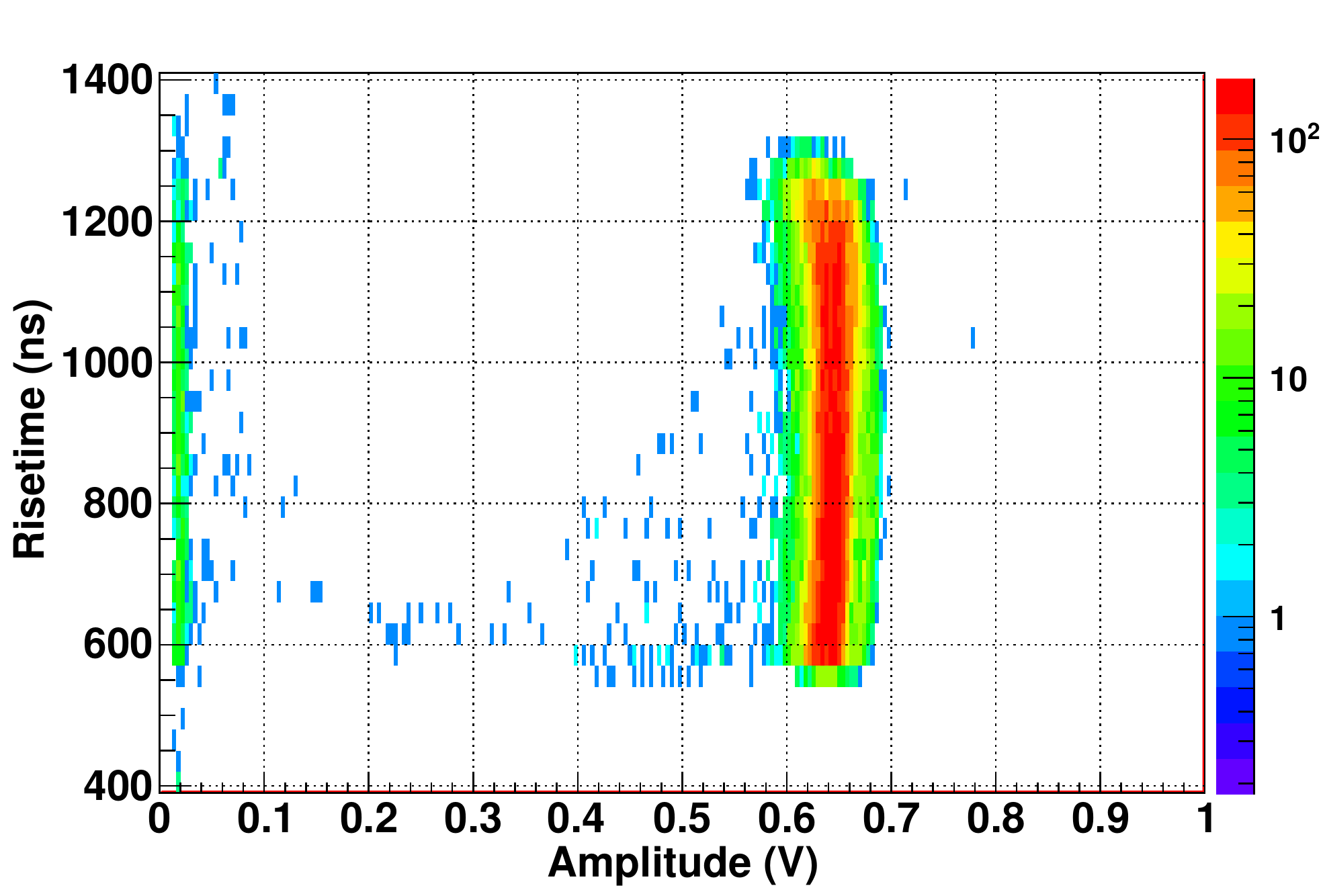}
\caption{ Left: Spectrum generated by the mesh pulses induced by the $^{241}$Am alphas, showing an energy resolution of 5.5\% FWHM. Right: Distribution of the risetime versus the amplitude of the mesh pulses induced by the $^{241}$Am alphas.}
\label{fig:SpecAmpRise}
\end{figure}

The spectra generated by the $^{241}$Am source showed an energy resolution of 5.5\% FWHM, as the one shown in figure \ref{fig:SpecAmpRise}. This value was independent of the drift voltage and the readout gain. To check the possible presence of attachment effects in the gas, mesh pulses were acquired by a LeCroy WR6050 oscilloscope. In an offline analysis, the amplitude and risetime of the pulses were calculated and the 2D distribution of these parameters was generated to look for correlations. Alpha events showed the same amplitude, independently of their risetime, i.e., their spatial direction. Therefore attachement effects are not observed. An example of these 2D distribution is shown in figure \ref{fig:SpecAmpRise} (right).

\begin{figure}[htb!]
\centering
\includegraphics[width=60mm]{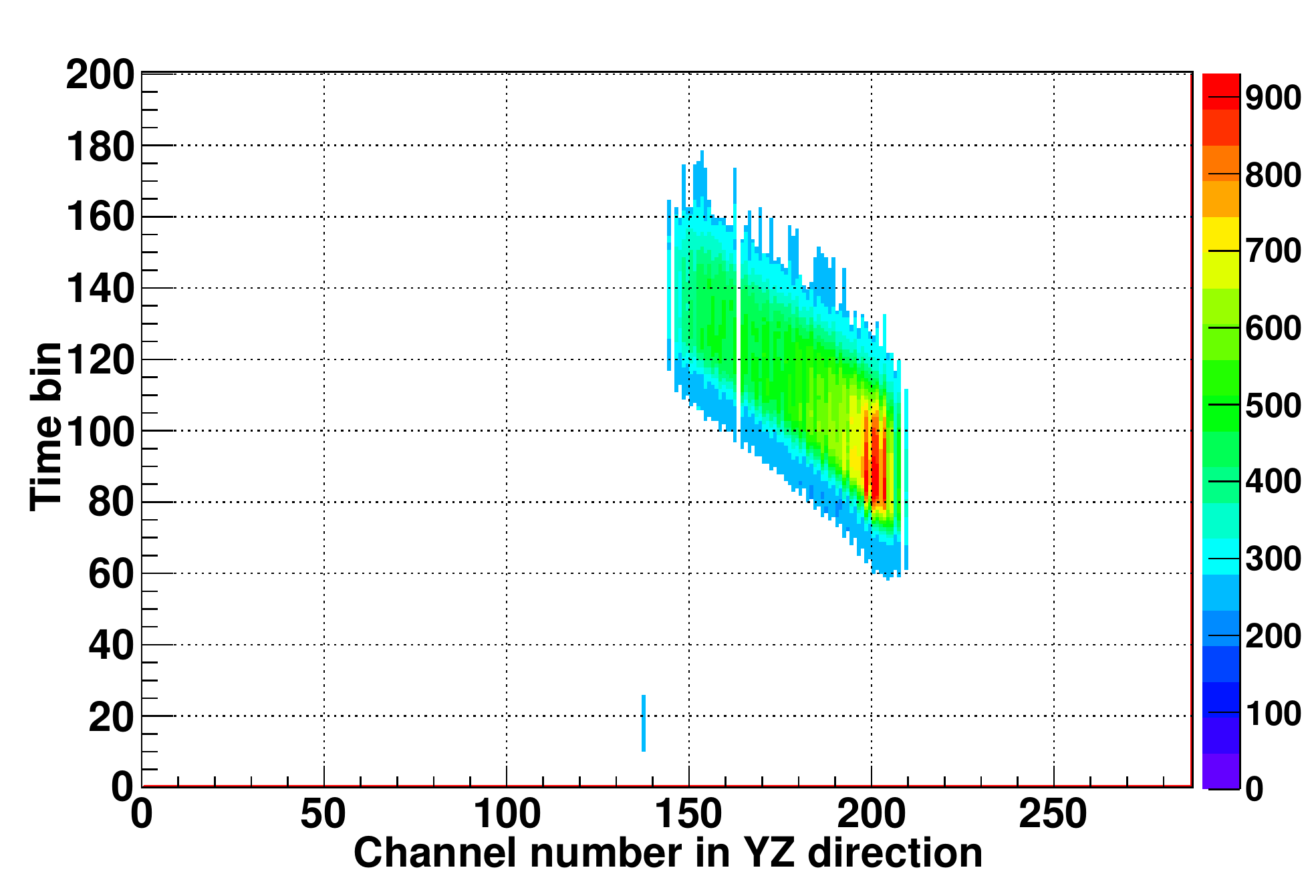}
\includegraphics[width=60mm]{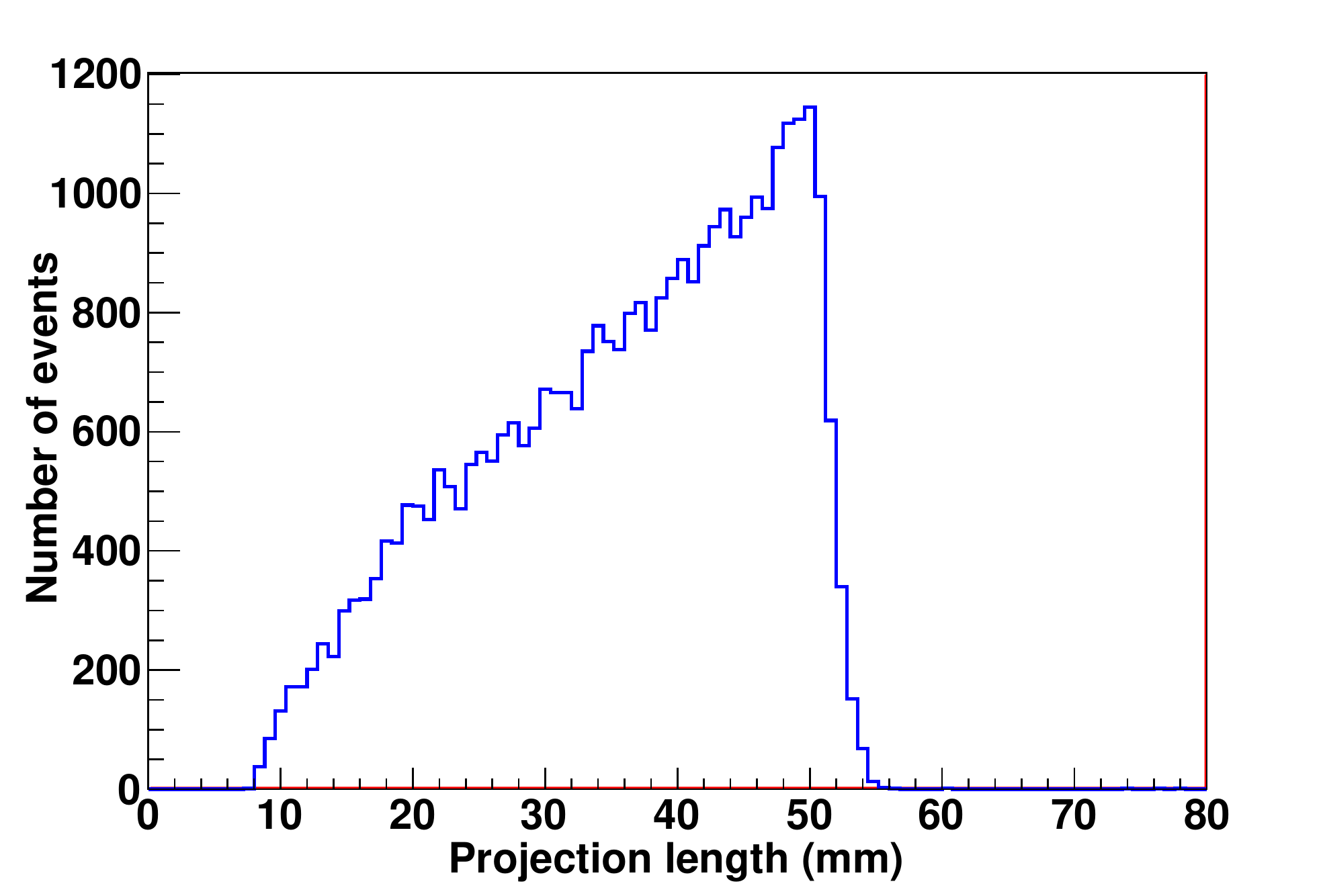}
\caption{ Left: An alpha event acquired by the T2K electronics. The track shows a final charge accumulation. Right: Distribution of the length of the track projection on the XY plane, calculated by the number of strips activated in each direction. The maximum length of the distribution matches the Geant4 simulation value (54\,mm).}
\label{fig:ProjLeng}
\end{figure}

Several alpha tracks were also acquired with the T2K electronics, as shown in figure \ref{fig:ProjLeng} (left). The mesh and drift voltages were respectively set to 400 and 820\,V, which correspond to a gain of 85 and a drift field of 70~V/cm. For each event, the length of the track projection on the XY plane was calculated and the distribution of this variable was generated. The maximum value obtained (54\,mm) matches the theoretical length expected for a 5.5 MeV\,alpha particles in an argon-based mixture (Iguaz \cite{Iguaz2010}).

The discrimination capabilities of this detector for low energy photons has also been  studied. Detailed results can be found in (Iguaz {\em et al.\/} \cite{IguazTIPP}).

\section{Conclusions and Outlook}

A bulk Micromegas detector has been designed and built for the MIMAC project. Characterisation measurements in the laboratory have shown good performance in terms of gain, uniformity, energy resolution and track measurements with  an Argon(95\%)-Isobutane(5\%) mixture at atmospheric pressure. The detector will be now tested with CF$_4$ in a neutron beam facility with neutrons of few keV using the specifically designed MIMAC electronics to reach the ultimate performance of the detector for the detection of Dark Matter. It is foreseen in the next coming months to test a module containing two 10 $\times$ 10\,cm$^2$ detectors in the Frejus underground laboratory. This measurement, in a realistic environment for a Dark Matter experiment, will give crucial information (background rejection as well as intrinsic contamination of the used materials) before the construction of a 1\,m$^3$ experiment. In any case the results obtained up to now validate the MIMAC concept for the construction of a large TPC for directional detection of Dark Matter.



\begin{thebibliography}{}

\bibitem[2010]{Ahlen2010}
Ahlen, S. et al., 2010, {\emph{International Journal of Modern Physics {\bf A 25}} 1-51}
[arXiv:0911.0323].

\bibitem[2009]{Allaoua2009}
Allaoua, A. et al, 2009, 
\emph{Radiat. Meas. \textbf{44}}  755 [arXiv:0812.0336] [physics.ins-det]. 

\bibitem[2010]{Andriamonje2010}
Andriamonje, S. et al, 2010,
\textit{JINST }\textbf{{5}} {P02001}.

\bibitem[2009]{Aune2009} 
Aune, S. et al., 2009,
\emph {J. Phys. Conf. Ser. {\bf 179}  012015}.

\bibitem[2009]{Baron2009}
Baron, P. et al.,  Presented at the Topical Workshop on Electronics for Particle Physics, Paris (France) 21/09/09--25/09/09.

\bibitem[2010]{Baron2010}
Baron, P. et al., 2010, {\emph{IEEE Trans. Nucl. Sci.} {\bf NS-57}}.

\bibitem[2010]{Billard12010} 
Billard, J. et al., 2010
\emph{Phys. Rev. D \textbf{82} } 055011. [arXiv:1006.3513] [astro-ph.CO].

\bibitem[2010]{Billard22010} 
Billard, J. et al., 2010,  \emph{Phys. Lett. \textbf{B 691}} 156-162, [arXiv:0911.4086] [astro-ph.CO]. 

\bibitem[2011]{Billard2011} 
Billard, J. et al., 2011,
 \emph{Phys. Rev. D \textbf{83}} 075002. [arXiv:1012.3960] [astro-ph.CO]. 

\bibitem[2010]{Bourrion2010}
Bourrion, O. et al., 2010,
{\emph{Nucl.\ Instrum.\ Meth. A} {\bf 662} 207},
[arXiv:1006.1335].

\bibitem[2009]{Dafni2009}
Dafni, T. et al., 2009,
{\emph{Nucl.Instrum.Meth.A} {\bf 60} 259-266}.
[arXiv:0906.0534 [physics.ins-det]]

\bibitem[2011]{Dafni2011}
Dafni, T. et al., 2011,
{\emph{Nucl.\ Instrum.\ Meth.A} {\bf 628}  172-176}.
 
\bibitem[2010]{Delbart2010}
Delbart, A. et al., 2010,
{\emph{Nucl.Instrum.Meth.A} {\bf 623} 105-107}.

\bibitem[1996]{Giomataris1996}
Giomataris, Y. et al., 1996,
{\emph{Nucl.\ Instrum.\ Meth. A} {\bf 376} 29-35}. 

\bibitem[2006]{Giomataris2006}
Giomataris, I. et al., 2006,
{\emph{Nucl.\ Instrum.\ Meth. A} {\bf 560} 405-408}
[physics/0501003].


\bibitem[2010]{Iguaz2010}
Iguaz, F.J.,  PhD thesis, Zaragoza, Universidad de Zaragoza, Zaragoza, Feb 2010. Presented: 25/02/2010. http://zaguan.unizar.es/record/5731

\bibitem[2011]{MimacSaclay}
Iguaz, F.J., et al., 
{2011} \textit{JINST }\textbf{{6}} {P07002}.

\bibitem[2011]{IguazTIPP}
Iguaz, F.J., et al., 2011,  Proceedings of the second international conference on Technology and Instrumentation in Particle Physics, Chigaco, United  States. To appear in Nuclear Instruments and Methods in Physics Research Section A. 

\bibitem[2010]{Richer2010}
Richer, J.P et al., 2010,
{\emph{Nucl.\ Instrum.\ Meth. A} {\bf 620} 470},
[arXiv:0912.0186].


\bibitem[2010]{Santos2010}
Santos, D. et al., 
[arXiv:0810.1137].

\bibitem[2010]{Santos2010} 
Santos, D. et al., 2011, \emph{J. Phys.  Conf.  Ser}., \textbf{309}, 012014 [arXiv:1102.3265].

\bibitem[2011]{Santos2011} 
Santos, D. et al.,   these proceedings. 


\bibitem[1988]{Spergel1988} 
Spergel, D. N., 1988, \emph{Phys. Rev. D \textbf{37}} 1353. 
    
\end{thebibliography}
\end{document}